# Water Isotope Separation using Deep Learning and a Catalytically Active Ultrathin Membrane

*Jinu. Jeong[1], Chenxing. Liang[2], and Narayana. R. Aluru[2,3] ***

*aluru@utexas.edu*

[1] Department of Mechanical Science and Engineering, The University of Illinois at Urbana−Champaign, Urbana, Illinois, 61801 United States

[2] Walker Department of Mechanical Engineering, The University of Texas at Austin, Austin, Texas, 78712 United States

[3] Oden Institute for Computational Engineering & Sciences, The University of Texas at Austin, Austin, Texas, 78712 United States

**ABSTRACT**

Water isotope separation, specifically separating heavy from light water, is a socially significant issue due to the usage of heavy water in applications such as nuclear magnetic resonance, nuclear power, and spectroscopy. Separation of heavy water from light water is difficult due to very similar physical and chemical properties between the isotopes. We show that a catalytically active ultrathin membrane (*e.g.*, a nanopore in $MoS_2$) can enable chemical exchange processes and physicochemical mechanisms that lead to efficient separation of deuterium from hydrogen, quantified as the $D_2O$ and deuterium separation ratio of 4.5 and 1.73, respectively. The separation process is inherently multiscale in nature with the shorter times representing chemical exchange processes and the longer timescales representing the transport phenomena. To bridge the timescales, we employ a deep learning methodology which uses short time scale ab-initio molecular dynamics data for training and extends the timescales to classical molecular dynamics regime to demonstrate isotope separation and reveal the underlying complex physicochemical processes.

**INTRODUCTION**

Heavy water, an isotope of $H_2O$, is of interest as many of its properties are quite similar to those of regular water. Heavy water is required for various applications, such as nuclear power plants,[1] spectroscopy,[2] and nuclear magnetic resonance.[3] Different production methods have been developed,[4,5] including chemical exchange,[6] distillation,[7] electrolysis,[8] crystallization,[9] and biological techniques;[10] however, only a few of these methods are used for industrial purposes owing to their relatively higher energy efficiency, capacity, and selectivity. The most popular method is the Girdler sulfide chemical exchange process. In the mixture of water and hydrogen sulfide, deuterium in the hydrogen sulfide is transferred to water and *vice versa* ($H_2O$ + HDS $\rightleftharpoons$ HDO + $H_2S$). The equilibrium constants depend on temperature (*e.g.*, 2.33 and 1.82 at 303 K and 403 K, respectively), which is utilized for concentrating deuterium in water.[4] Distillation exploits the boiling temperature of heavy water being higher than regular water. Water vapor is enriched with regular hydrogen, and deuterium is concentrated in liquid water. Electrolysis decomposes water into hydrogen (deuterium) and oxygen gases, where regular hydrogen is more likely to be decomposed than deuterium. However, these methods are expensive, and advances in separation technologies are needed to tackle isotope separations.

Owing to the recent advances in nanotechnology, promising alternative approaches have been developed. Mohammadi *et al*. experimentally studied the physicochemical properties of graphene oxide membranes that affect heavy-water filtration performances and discovered that a low oxidation level of membranes gives rise to a high heavy-water rejection rate.[11] Furthermore, Ono *et al*. suggested activated carbon fiber (ACF) as a heavy water separation medium by generating adsorption and desorption differences controlled by external pressure.[12]

In this work, we investigate separation of water isotopes using ultrathin membranes with catalytic sites. While there are a number of ultrathin membranes, including two-dimensional materials and MXenes, that contain catalytic sites, as a model ultrathin membrane we study $MoS_2$. Owing to its selective transport characteristics, $MoS_2$ nanopore is widely studied for molecule separation research, such as water desalination,[13] Deoxyribonucleic Acid (DNA) sequencing,[14] and molecule separation.[15] A single layer $MoS_2$ has good mechanical properties with an effective Young's modulus of 270 ± 100 GPa,[16] exhibiting high structural stability. The presence of Mo atoms in the pore region can facilitate chemical exchange over atomic length scales, and we exploit this phenomenon for separation for water isotopes.

The physicochemical processes involved in water isotope separation include adsorption/desorption, molecule dissociation and formation of new molecules, chemical reactions, induced driving forces, and transport. The timescales associated with these processes can range from femtoseconds to nanoseconds and longer. Ab-initio molecular dynamics (AIMD) calculations, combining density functional theory with classical molecular dynamics, can reasonably describe the physicochemical processes involved in water isotope separation. However, these calculations are limited to short timescales – typically, several tens of picoseconds. As a result, it is difficult to observe long timescale physicochemical processes such as complete translocation of molecules through the nanopore. In this work, to overcome this limitation, we employ a deep learning methodology which uses AIMD data for training. We demonstrate that the deep learning methodology is able to resolve the physicochemical processes involved in water isotope separation processes.

**METHODS**

We conducted a quantum-level computational study on heavy water separation through a $MoS_2$ nanopore. A quantum-accurate method is required to explore the complete physicochemical transport[17,18] and the catalytic effect of the Mo atom on water.[19–21] The AIMD simulation is one of the popular and quantum-accurate approaches in molecular simulation. The strength of AIMD simulation is its quantum-level precision that can help study reactive systems;[22–25] however, the method is impractical for large systems owing to massive computational costs. To reduce the computational cost of simulating reactive systems, reactive interatomic potentials based on quantum-accurate methods have been developed.[26,27] Neural network potentials[28–34] are especially popular owing to their universal reproducibility for a target system without ad hoc approximation and fitting functions. Here, we train neural network potentials for AIMD data to generate long-time reactive simulation trajectories, capturing physical and chemical details in the separation process.

**AIMD simulation**

AIMD simulations are used to solve Newtonian dynamics where the force is obtained from the density functional theory.[35] We use the Vienna ab initio simulation (VASP) package, in which the projector-augmented wave method with a plane-wave basis set is used.[36,37] The Perdew–Burke–Ernzerhof (PBE) exchange-correlation functional[38] from generalized gradient approximation is employed, and the DFT-D3 method[39] is used for van der Waals interaction correction. Energy cutoff, Gaussian smearing width, and timestep size are taken to be 400 eV, 0.05 eV, and 0.5 fs, respectively. No external pressure is applied, and the temperature is maintained at 400 K. We perform AIMD simulations, having the same amount of $H_2O$ and $D_2O$ (81 for each) in the system.

In this study, the MoS$_2$ nanopore is constructed using a repeating arrangement of molybdenum and sulfur atoms, referred to as the mixed-type MoS$_2$ nanopore[13]. This nanopore is positioned at the system's center and is created by extracting 12 Mo atoms and 24 S atoms from a MoS$_2$ membrane. It possesses an approximate diameter of 13 Å, allowing room for molecules pass through. The masses of Mo, S, O, H, and D atoms are 95.94, 32.066, 16, 1.013, and 2.013 Da, respectively.

Because the computational cost of running AIMD simulations for the target system is considerable, it is useful to generate data for the subsystems containing essential physics and use the data for pretraining. This is followed by fine-tuning where the pre-trained neural network is further trained with the whole system data. We consider three subsystems designed to highlight water–water interaction in bulk, water–MoS$_2$ wall interaction, and water–pore interaction (Supporting Information S.2 for details).

**DeePMD training**

Using the AIMD data generated above, we train a DeePMD-SE[31] model, where a neural network is trained to predict the potential energy for a descriptor containing information about atomic configuration in a local neighborhood. The descriptor is an embedded matrix that takes generalized local coordinates as input. The local coordinates of atom $i$, $R^i$, comprise $N_i$ row vectors, $\{x_{ji}, y_{ji}, z_{ji}\}$, which are then transformed to global coordinates $\tilde{R}^i$ the row vector of which can be written as $\{s(r_{ij}), \hat{x}_{ji}, \hat{y}_{ji}, \hat{z}_{ji}\}$. Here, $s(r_{ij})$ is a differentiable scalar weighting function that is defined as

$$s(r_{ij}) = \begin{cases} \dfrac{1}{r_{ji}} & r_{ji} < r_{cs} \\ \dfrac{1}{r_{ji}} (\dfrac{1}{2}\cos\left(\pi\dfrac{r_{ji} - r_{cs}}{r_c - r_{cs}}\right) + \dfrac{1}{2}) & r_{cs} \leq r_{ji} < r_c' \\ 0 & r_c \leq r_{ji} \end{cases} \qquad (2)$$

where $r_{cs}$ and $r_c$ signify smooth cutoff and cutoff distances, which are 6.00 and 7.00 Å, respectively. Additionally, $\hat{x}_{ji}$, $\hat{y}_{ji}$, and $\hat{z}_{ji}$ are obtained by multiplying $s(r_{ij})/r_{ji}$ with $x_{ji}$, $y_{ji}$, and $z_{ji}$. The embedding network $G\left(s(r_{ji})\right)$ is used to map $s(r_{ji})$ to $M_1$ outputs, i.e., $G: \mathbb{R}^1 \mapsto \mathbb{R}^{M_1}$, where we set $M_1$ as 64. This neural network maps a single value through several hidden layers, to the output layer on embedding feature dimension. Its difference to fitting layer is that embedding networks are atom-type specific and independent from each other and located at the head of the whole neural network. The fitting network, the tail part, takes the output of the embedding networks, and all types of atoms share the same neural network. Having $k$th number of $s(r_{ji})$ inputs, one can obtain $\left(\mathcal{G}^i\right)_{jk} = \left(G\left(s(r_{ji})\right)\right)_k$. Finally, the feature matrix $\mathcal{D}^i$ can be obtained by computing $\left(\mathcal{G}^{i1}\right)^T \tilde{R}^i \left(\tilde{R}^i\right)^T \left(\mathcal{G}^{i2}\right)$, which is fed into the fitting neural network for potential energy prediction. The fitting neural network or fitting net can be expressed as

$$E = \sum_i^N NN_s(\mathcal{D}^i) \qquad (3\text{-}1)$$

$$NN_s(\cdot) = \mathcal{L}^{out}\left(\mathcal{L}^4\left(\mathcal{L}^3\left(\mathcal{L}^2(\mathcal{L}^1(\cdot))\right)\right)\right) \qquad (3\text{-}2)$$

where $E$, $N$, and $NN_s$ represent the total potential energy, the total number of atoms, and the feed-forward fitting net model, respectively. The fitting net comprises four fully connected layers containing 128 nodes followed by the tanh activation function ($\mathcal{L}(\cdot)$) and output layers $\mathcal{L}^{out}(\cdot)$. The loss function is defined as

$$L(p_\epsilon, p_f, p_\xi) = \frac{1}{|\mathcal{B}|} \sum_{l \in \mathcal{B}} p_\epsilon \left| E_l^{(GT)} - E_l^{(pred)} \right|^2 + p_f \left| F_l^{(GT)} - F_l^{(pred)} \right|^2 \quad (4)$$

$$+ p_\xi \left| \Xi_l^{(GT)} - \Xi_l^{(pred)} \right|^2$$

where $\mathcal{B}$ symbolizes minibatch and $|\mathcal{B}|$ stands for the minibatch size, which is two in our training procedure. $E_l^{(GT)}$, $F_l^{(GT)}$, and $\Xi_l^{(GT)}$ refer to the potential energy, force, and virial of the system, which can be obtained using AIMD simulation. $E_l^{(pred)}$, $F_l^{(pred)}$, and $\Xi_l^{(pred)}$ indicate the potential energy, force, and virial of the system, computed using the DeePMD model. $p_\epsilon, p_f$, and $p_\xi$ are tunable prefactors that prioritize or underestimate each component in the loss function. Moreover, $p_\epsilon$ and $p_f$ have 0.02 and 1000 for initial values, which gradually reach unity. We assign zero for $p_\xi$ during the entire training process, meaning our neural network is mainly trained to prioritize potential energy and force reproducibility. Using the ADAM optimizer with an initial learning rate of $1e^{-3}$ decaying until $3.51e^{-8}$, neural network training is iterated for 1,000,000 steps. The neural network is initially trained with subsystem data (see Supporting Information S.1. for details) and fine-tuned using the entire system data. The fine-tuned neural network has an energy error of $4.05 \times 10^{-3}$ eV and a force error of $1.84 \times 10^{-2}$ eV/Å. For all of the above training procedures, we use DeePMD-kit[30] written in Tensorflow.[40] Further validation considering physical data reproducibility are discussed in the Supporting Information S.3. section.

**DeePMD simulation**

LAMMPS,[41,42] included in the DeePMD-kit,[30] is used to perform time integration and to generate trajectory in a DeePMD simulation. Force is computed by obtaining the negative potential energy gradient with respect to atomic coordinates. We assign the same atomic masses used in AIMD analysis for the DeePMD simulation. The Nosé–Hoover thermostat[43] with a time constant of 100 fs is applied, maintaining the system temperature at 400 K.

**RESULTS**

Heavy-water separation is studied using a neural network potential-based molecular dynamics simulation for a single layer $MoS_2$ nanopore equilibrated with 81 $H_2O$ and 81 $D_2O$ molecules. We observe active chemical reactions in confinement, reactive translocation of molecules, and water isotope separation phenomena in our system (Figure 1.A). A molecule in and near the pore undergoes various chemical reactions (Figure 1.B), such as water molecule adsorption to Mo atom (Figure 1.B.i), hydrogen detachment from Mo-attached water molecule (Figure 1.B.ii), proton donation to a water molecule to form a hydronium molecule (Figure 1.B.iii), proton detachment to from a water molecule (Figure 1.B.iv), and proton donation to a hydroxyl group (Figure 1.B.v). To clarify the molecular forms of OH and OD, we use OH/OD attachment (to Mo atom), OH/OD bond (as a part of a water molecule), and $OH^-/OD^-$ ions (solvated in water). We note that these reactions can also occur for deuterium and deuterium-containing molecules. Continuous proton donation and detachment can change the isotopic composition of a water isotope. To study heavy water separation performance, we classify pore physics into inflow, backflow, and translocation (Figure 1.C), which denote the entry of a water isotope molecule into the pore, return of a molecule to the originating reservoir, and escape of the water molecule to the opposite reservoir.

Figure 2.A shows the time history of each translocation event. We track the translocating molecules for 2ns of simulation trajectories and observe that about 10 of the 23 translocated water isotopes eventually escaped the pore exhibiting a different isotopic state. In Figure 2.B, the number of translocations for each water isotope group is counted, where the translocations are highest for HDO, $D_2O$, and $H_2O$. Similarly, the number of hydrogen and deuterium atom translocations is visualized in Figure 2.C, where deuterium translocation is higher than hydrogen translocation by 16. This provides another indication of the separation efficiency, as the fractions of $H_2O$, HDO, and $D_2O$ after the separation are determined from the number of translocated H and D atoms. The translocation of any hydrogen (or deuterium) atom is counted regardless of the molecular form. For example, the translocation of the H atom includes hydrogen translocation in $H_2O$, HDO, $H^+$, $H_3O^+$, $H_2DO^+$, and $HD_2O^+$. Separation performance is quantified with separation ratios of $D_2O$ over $H_2O$ and deuterium over hydrogen, which are 4.5 and 1.73, respectively. The deuterium over hydrogen separation ratio is smaller than that of $D_2O$ over $H_2O$ as the translocation of HDO, the molecule having the highest degeneracy and population, contributes to both hydrogen and deuterium translocations. Due to the catalytic site in the pore and generation of free $H^+/D^+$ ions, chemical reactions are active in the pore (Figure 2.D); thus, the translocating molecules undergo chemical reactions, changing their isotopic composition giving rise to reactive translocation. For example, in Figure 2.E, $H_2O$ enters the pore, undergoes chemical reactions, and translocates to the opposite reservoir. Similarly, in Figure 2.F, a $D^+$ ion detached from $D_3O^+$ enters the pore and translocates with a different oxygen atom. We further study the physicochemical mechanisms that contribute to the dominance of $D_2O$ and deuterium atom during the translocation, *i.e.*, Mo-attached molecules and the composition of net inflow that is defined as the number of each molecule in the inflow subtracted by that of backflow (see Figure 1.C for details). Considering three molecule types (water isotope group, hydronium isotope group, and $H^+/D^+$ ion group), we observe that a higher number of deuterium-dominant molecules are translocated than their counterparts in the water isotope group and the hydronium isotope group, while more $H^+$ ions are translocated than $D^+$ ions (Figure 3.A, see Supporting Information S.4. section and Figure S.4 for the inflow and backflow information). The information regarding the composition of net inflow is crucial, as it allows us to distinguish the contributions of non-reactive diffusion mechanism (involving water and hydronium isotopes, Figure 3.B left) from those of the Grotthuss mechanism (involving hydrogen ion isotopes, Figure 3.B center and right), which cannot be investigated through translocation data (Figure 2. A-C). We observe that there are more hydrogen ions than deuterium ions in the net flow; however, the net inflow of other molecules is governed by deuterium-dominant water and hydronium isotopes. This results in a higher translocation of deuterium atoms compared to hydrogen. In the translation-driven inflow (Figure 3.B left), a molecule crosses the boundary without chemical reaction, and water isotope and hydronium isotope molecules rely on this mechanism. In Grotthuss mechanism-driven inflow, $H^+/D^+$ is detached from a hydronium isotope and donated to a nearby water isotope molecule to form hydronium isotope with a different oxygen atom (Figure 3.B center). It is also plausible that a free $H^+/D^+$ ion directly combines with a hydroxyl group to form a water isotope (Figure 3.B right), and similarly, a free $H^+/D^+$ ion detached from nearby hydronium isotope molecule can be donated to a hydroxyl group. We also study the isotopic composition (Figure 3.C) and physical behavior of a hydroxyl group (Figure 3.D), since they are closely related to chemical reactions in the pore, affecting reactive translocation and isotope separation performance. OH attachment is preferred over OD attachment, which is quantified with the OH to OD attachment time ratio, where a value of 1.25 is obtained. Another interesting observation is that the oxygen atoms attached to Mo are almost stationary and do not escape the pore (Figure 3.D). This indicates that when a translocating hydrogen isotope combines with the O atom in a hydroxyl group, it will stay next to the oxygen, *i.e.*, at the center of the pore, until another $H^+/D^+$ ion combines with the hydroxyl group. The OH/OD attachments act like a hindrance to hydrogen isotope translocation, and the H atom is more affected by this physicochemistry. As a result, H atoms are less translocated than that of D atoms. While it is counterintuitive that more deuterium based molecules are transported over hydrogen based molecules, similar phenomena have been observed in other works. Oh *et al*. demonstrated that $D_2$ could be separated using a nanopore with pyridine molecules as a size-adjustable aperture.[44] They reported that $H_2$ is larger than $D_2$, originating from the nuclear quantum effect. Niimura *et al*. conducted $D_2$ gas separation studies considering various nanoporous materials (*e.g.*, ACF, carbon molecular sieve, zeolite, and carbon nanotubes) and observed more $D_2$ being transported.[45]

To understand the underlying physics, we further study the energy differences between $H_2O$ and $D_2O$, OH/OD bond stability, and $H^+/D^+$ ion transfer time. Figure 4.A shows conceptual potential energy surfaces (PES) of OH and OD bonds, as well as the physical quantities related to the stability of OH/OD bond: average potential energies of OH/OD bonds, the fluctuations of OH/OD bond, the fluctuations of force acting on H/D atom. In general, OD bond is more stable than OH bond since the heavier mass of deuterium reduces the fluctuation of OD bond, which reduces the potential energy of OD bond and increases the dissociation energy of OD bond. We believe there is a contribution from quantum mechanical phenomena (e.g. kinetic isotope effect), on the top of classical chemistry (e.g. reaction rate difference). We analyze the potential energy differences between $H_2O$ and $D_2O$, in bulk and confined environments to obtain physical insights. We perform ab initio molecular dynamics (AIMD) simulations for a bulk system comprising 32 $H_2O$ molecules at 400 K and a separate bulk system comprising 32 $D_2O$ molecules at 400 K (Supporting information S.1. and Figure S.1.A top). Similarly, AIMD simulations were performed for a $MoS_2$ nanopore system with four $H_2O$ molecules confined in the pore at 400 K and a separate system with four $D_2O$ molecules confined in the pore at 400 K (Supporting information S.1. and Figure S.1.A bottom). The total energies of bulk $H_2O$, bulk $D_2O$, confined $H_2O$, and confined $D_2O$ are

−14.454 eV, −14.4508 eV, −580.345 eV, and −580.706 eV, respectively. Note that they have the same averaged kinetic energy since the systems are equilibrated at the same temperature. To systematically compare the preference between the water isotopes depending on their environment, we compute the energy differences between isotopic systems and normalize by the number of substituted water isotopes: $\Delta\varepsilon^{bulk} = \left(E_{H_2O}^{bulk} - E_{D_2O}^{bulk}\right)/32$ and $\Delta\varepsilon^{conf} = \left(E_{H_2O}^{conf} - E_{D_2O}^{conf}\right)/4$, obtaining $-1.01 \times 10^{-5}$ eV and $9.02 \times 10^{-2}$ eV, respectively. We present the magnitude of those values in Figure 4B. In order to evaluate the significance of the values, we compare them with the magnitude of $H_2O$ total energy ($4.52 \times 10^{-1}$ eV, the blue horizontal line in Figure 4.B) and the standard deviation of $H_2O$ kinetic energy ($1.81 \times 10^{-2}$ eV, the red horizontal line in Figure 4.B). The magnitudes of $\Delta\varepsilon^{bulk}$ and $\Delta\varepsilon^{conf}$ are 0.0223% and 20.0% of the magnitude of $H_2O$ energy, respectively, signifying that $H_2O$ and $D_2O$ are almost equally preferred in bulk, but $D_2O$ is more preferred in nanopore, under the effect of catalytic effect of $MoS_2$ nanopore edge. Also, the magnitudes of $\Delta\varepsilon^{bulk}$ and $\Delta\varepsilon^{conf}$ have different order from that of the standard deviation of $H_2O$ kinetic energy: 0.0559% and 498%, which supports their difference is not caused by thermal energy fluctuation.

We also analyze the fluctuation of the force acting on H/D and the fluctuation of OH/OD bond length, which are different quantities related to the stability of OH/OD bond. As the mass of deuterium is heavier than that of hydrogen, it has a smaller vibration amplitude. This reduces deuterium deviating from the position of zero net force ($r^*$ in Figure 4.A), stabilizing the molecule and elevating the activation energy required for chemical reaction. The deviation of atom position from $r^*$ can be quantified by the fluctuation of force acting on hydrogen (deuterium) atom and the fluctuation of OH/OD bond length. Herein, we use the standard deviation of the force magnitude acting on hydrogen (or deuterium) atoms and the standard deviation of OH (or OD) bond length, respectively. We perform AIMD simulations for the two systems: 12 $H_2O$ molecules that can freely enter and escape the pore confined by an unmovable water shell (Supporting information S.1. and Figure S.1.B left) and the same system substituting freely movable $H_2O$ with $D_2O$ (Supporting information S.1. and Figure S.1.B right). As shown in Figure 4.C., the computed force magnitude fluctuation of $H_2O$ outside- and inside-pore are 0.889 eV/Å and 1.56 eV/Å, respectively, and the counterparts of $D_2O$ are 0.835 and 1.03 eV/Å. In bulk, the hydrogen (deuterium) force fluctuations of water isotopes are almost the same, but it is significantly increased for $H_2O$ in confinement. These results reveal that $H_2O$ in bulk is chemically as stable as $D_2O$ in bulk, while $H_2O$ inside the pore is chemically less stable than $D_2O$ inside the pore, because of the catalytic effect of $MoS_2$ on water isotopes. Similar results are observed in the bond length fluctuation analysis. The calculated force magnitude fluctuation of $H_2O$ outside- and inside-pore are 0.255 and 0.322 Å, respectively, and the corresponding fluctuations in force for $D_2O$ are 0.217 and 0.289 Å (Figure 4.D). The bond length fluctuations are of the highest for $D_2O$ outside the pore, $H_2O$ outside the pore, $D_2O$ inside the pore, and $H_2O$ inside the pore, respectively, consistent with the force magnitude analysis. The AIMD analyses on energy, hydrogen (deuterium) force fluctuation, and OH/OD bond length fluctuation indicate that an OH bond in the pore is relatively less stable compared to an OD bond in the pore. The molecules with rich OD bonds ($D_2O$, $D_3O^+$, and $HD_2O^+$) are less reactive, and $H^+$ ion is more prevalent in the pore. A similar phenomenon is reported by Pasquini *et al*., where catalytic effect of Co-based amorphous-oxide catalysts on water is decreased by 82% for $D_2O$.[46] The more reactive OH bond increases the probability of $H^+$ ions becoming trapped in Mo-active sites, and the molecules with rich OD bonds are favored, resulting in a lower total energy of the system.

To understand proton transfer kinetics within the nanopore, we consider an oxygen atom attached to the Mo-site and consider transfer of proton/deuterium from a $H_3O^+$/ $D_3O^+$ ion inside the pore. AIMD simulations are performed for a $H_3O^+$ ion at various distances ($d_{OO}$) with an oxygen atom attached to the Mo-site of the nanopore (Supporting information S.1 and Figure S.1.C left) and for an identical system with hydrogen substituted by deuterium (Supporting information S.1 and Figure S.1.C right). We compute transfer time, which is defined as the duration between the discharge of a proton isotope hydronium isotope and its subsequent attachment to the oxygen atom bound to the Mo atom; it is observed that the $H^+$ transfer time is shorter than the $D^+$ transfer time (Figure 4.E) because of the higher mobility of $H^+$ ions and lower stability of the OH bond. Here, the mobility of each ion plays an important role as once the attached oxygen at Mo-site receives one of the $H^+$ or $D^+$ ions, it is less likely to accommodate more $H^+/D^+$ ions. This first-come-first-serve mechanism, driven by mobility difference, is reported by Weinrauch *et al.*, where $H_2$ adsorption can dominate over $D_2$.[47] This corresponds to the higher occupancy time of OH attachment than that of OD attachment, shown in Figure 3.C. Higher occupation of OH attachment can be also explained by collision rate theory. The rate equation in collision theory, $r(t) = Z\rho e^{-\frac{E_a}{RT}}$ explains that the reaction frequency is proportional to the collision frequency ($Z$), steric factor ($\rho$), and the Boltzmann factor ($e^{-\frac{E_a}{RT}}$). In this context, the collision frequencies for the hydrogen and deuterium systems are the only difference since hydrogen atoms are lighter than their counterparts, resulting in higher average velocities and increased collision probabilities. Similar phenomenon is observed by Phillips, where the chemical reaction of $CH_3^+$ with $H_2$ is more active than and that of $D_2$.

**CONCLUSION**

We showed that a catalytically active $MoS_2$ nanopore can effectively separate $D_2O$ and deuterium with a separation ratio of 4.5 and 1.73, respectively. To understand the underlying mechanism, we analyze the dominant molecules comprising the net inflow and their transport mechanisms. $D_2O$, D-dominant hydronium isotopes, and $H^+$ ions constitute the net inflow compared to their isotopic counterparts. Translation-driven inflow mechanism governs the net inflow of water and hydronium isotopes that contain oxygen atoms, and the proton-jumping mechanism governs the net inflow of $H^+$ and $D^+$ ions. Furthermore, we observed that OH attachment to a Mo-active site is preferred to OD attachment and the attached oxygen atoms are almost stationary, temporarily trapping a hydrogen isotope.

To gain physical insights, we analyzed the fluctuation of the force acting on hydrogen isotopes and the fluctuations of OH (OD) bond length, finding that the OH bond is less stable than the OD bond in the pore, while they have nearly identical stability in bulk. A similar tendency is observed in energy analysis: the energy difference between $H_2O$ and $D_2O$ in bulk is negligible, and that in confinement is considerable. These analyses account for the high number of $D_2O$ and D-dominant hydronium ions in the net inflow. The analysis also explains the dominance of $H^+$ ions in the net inflow and the preference for OH attachment to Mo atoms. The higher fluctuation in OH bond length increases the chance of proton detachment, and the lower mass of a hydrogen atom causes higher mobility of $H^+$ ions. These effects promote more frequent attachment and detachment of proton to an oxygen atom in a hydroxyl group. The separation mechanism and physical interpretation presented here can be exploited for other separation mechanisms.

**SUPPORTING INFORMATION**

Additional information regarding the comparison of stability between $H_2O$ and $D_2O$, the configurations of system used for DeePMD training data generation, and the physics reproducibility test of DeePMD simulations is available.

**ACKNOWLEDGMENT**

The work on water was supported by the Center for Enhanced Nanofluidic Transport (CENT), an Energy Frontier Research Center funded by the U.S. Department of Energy, Office of Science, Basic Energy Sciences under Award No. DE-SC0019112. All other aspects of the work are supported by the National Science Foundation under grants 2140225 and 2137157. The authors also acknowledge the Texas Advanced Computing Center (TACC) at The University of Texas at Austin for providing the computing resources on Stampede2, Frontera, and Lonestar6 under Allocation Nos. TG-CDA100010, DMR20002, and DMR22008, respectively. Additionally, the authors acknowledge the use of the Illinois Campus Cluster Program (ICCP) in conjunction with the National Center for Supercomputing Applications (NCSA).

**TOC IMAGES**

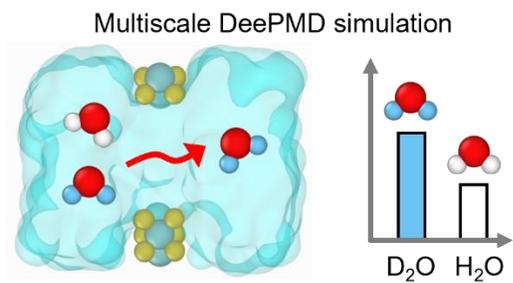

# FIGURES

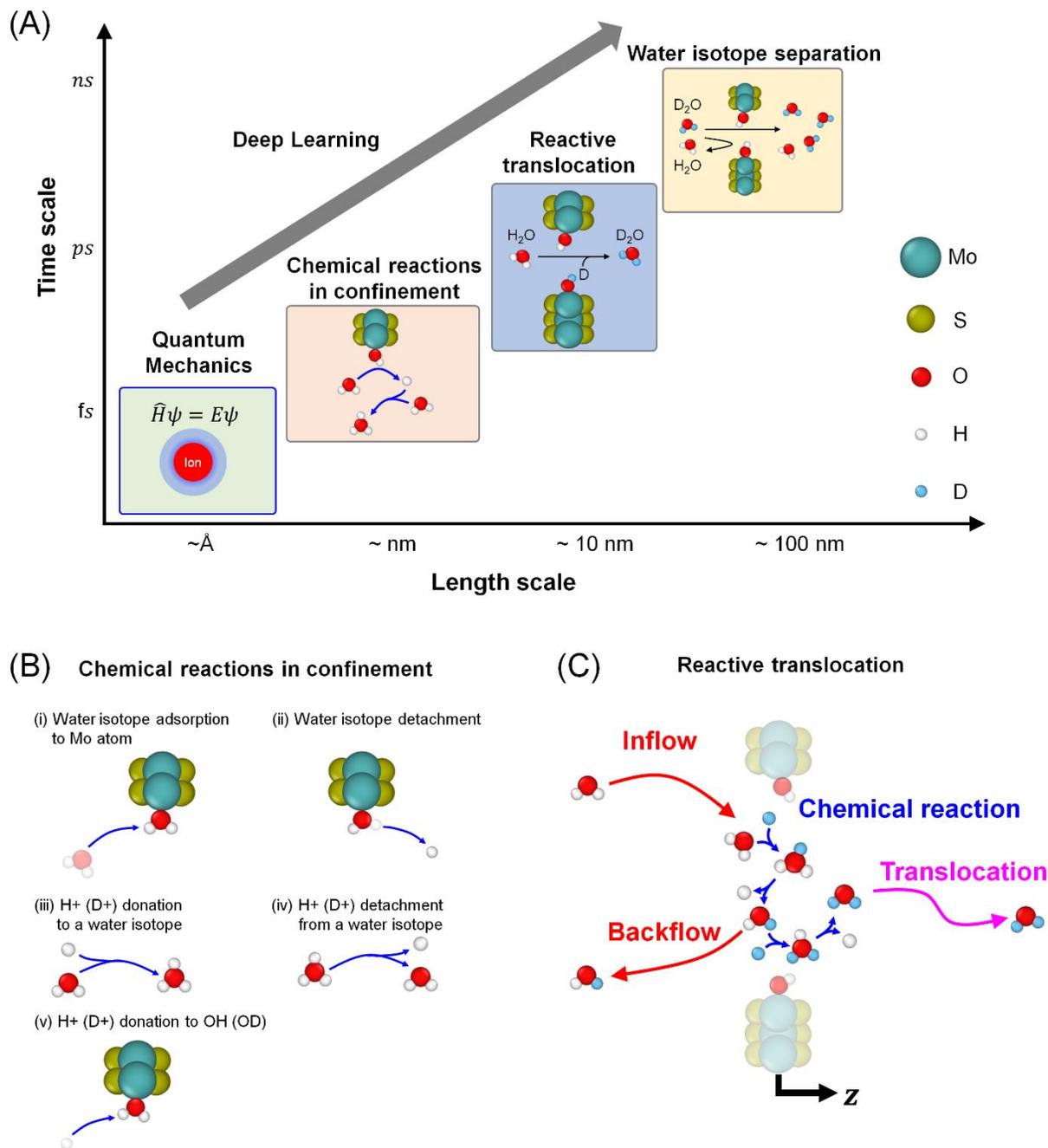

**Figure 1.** (A) The physical and chemical phenomena involved in water isotope separation using a $MoS_2$ nanopore. The deep learning approach is used to efficiently link the physicochemical phenomena ranging from chemical reactions to long-time dynamics. (B) Representative chemical reactions observed in the pore. While the schematics are shown using only the regular hydrogen, these reactions can occur for deuterium and a deuterium-containing molecule. (i) Adsorption of a water isotope to the Mo atom in the pore. (ii) Detachment of a proton (or a deuterium) from the water molecule attached to the Mo atom. An $OH^-$ group is now attached to the Mo atom. (iii) The detached $H^+$ or $D^+$ ion (from step (ii)) combines with a water isotope molecule to form a hydronium ion isotope. (iv) A hydronium ion isotope has lower stability and one of the hydrogen isotopes detaches and is available as a free charge. The freely available $H^+$ or $D^+$ can combine with a nearby water isotope as in step (iii). (v) Freely available $H^+$ or $D^+$ ions can combine with the hydroxyl group on the Mo atom to form a water isotope molecule followed by step (ii). (C) Schematic illustrating a molecule undergoing reactive translocation. In this example, an $H_2O$ molecule enters the pore, undergoes $H^+/D^+$ attachment/detachment processes and leaves the pore as a $D_2O$ molecule. The schematic also shows backflow (from the pore to the reservoir) of molecules, which can be different from the entering molecules during the chemical exchange within the pore.

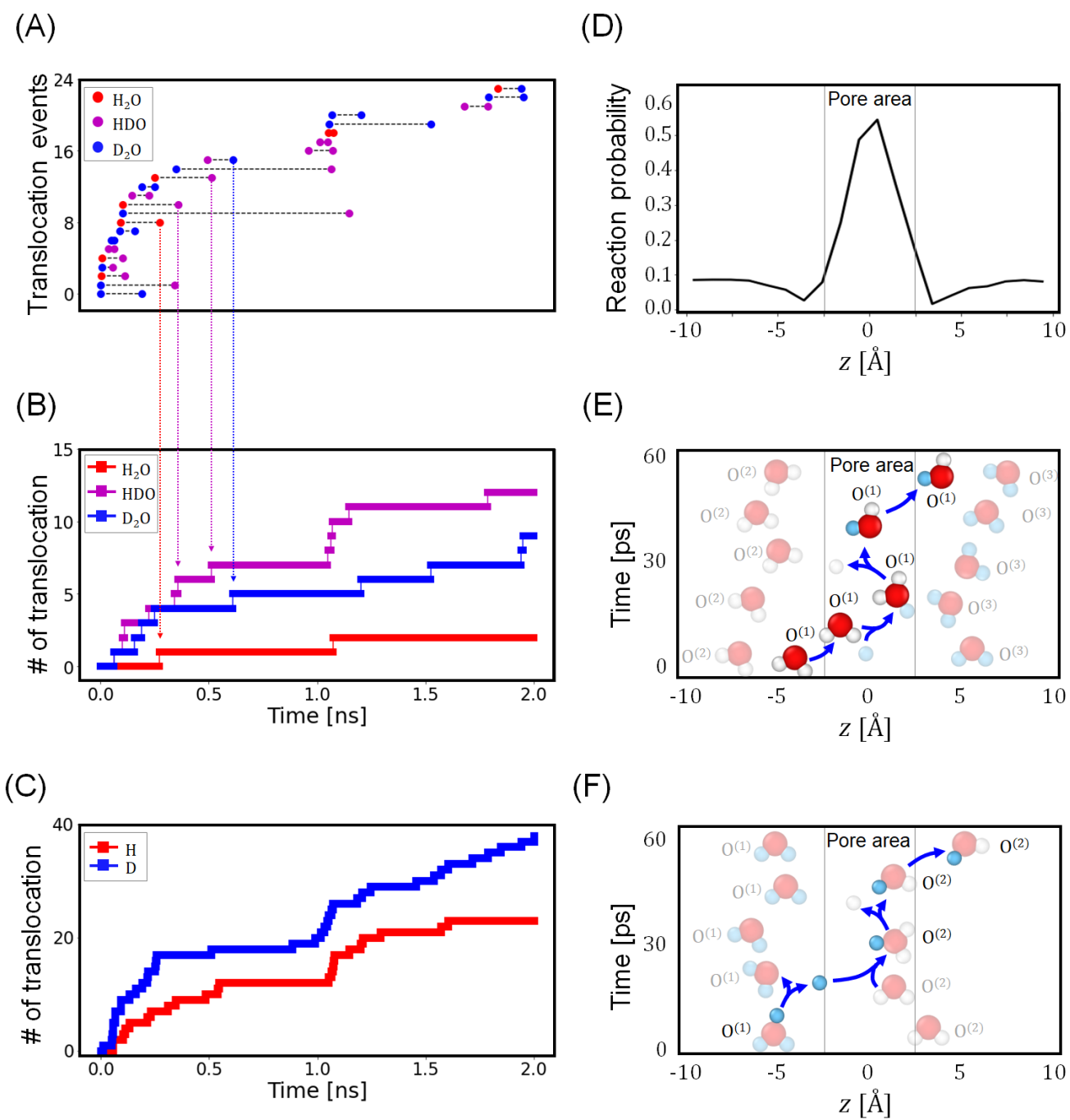

**Figure 2.** (A) The history of translocation events. A translocating water molecule is denoted by two circles connected by a dashed line, where the color of each circle indicates the isotopic state at the pore entrance (beginning time) and exit (end time with molecule leaving the pore), respectively. (B) The number of translocations of each molecule visualized over time. (C) The number of translocations of the H and D atoms visualized over time. (D) The probability of reaction with respect to the z-direction. The chemical reactions are enhanced as a molecule enters the pore. Chemical exchanges in the pore lead to a translocating molecule change its isotopic composition. (E) An example of a reactive translocation trajectory with chemical exchange – $H_2O$

enters the pore and combines with a $D^+$ ion, forming $H_2DO^+$. Since the product is unstable, one of the hydrogen isotopes detaches, which is regular hydrogen in this example, forming HDO. The molecule then escapes the pore as an HDO molecule. Notice that the oxygen atom is the same in this translocation. (F) An example of a hydrogen isotope translocation. Deuterium discharged from $D_3O^+$ enters the pore and combines with $H_2O$, forming $H_2DO^+$. The product is unstable and splits into $H^+$ and HDO. The HDO escapes the pore. This is an example of a hydrogen isotope translocation with a different oxygen atom.

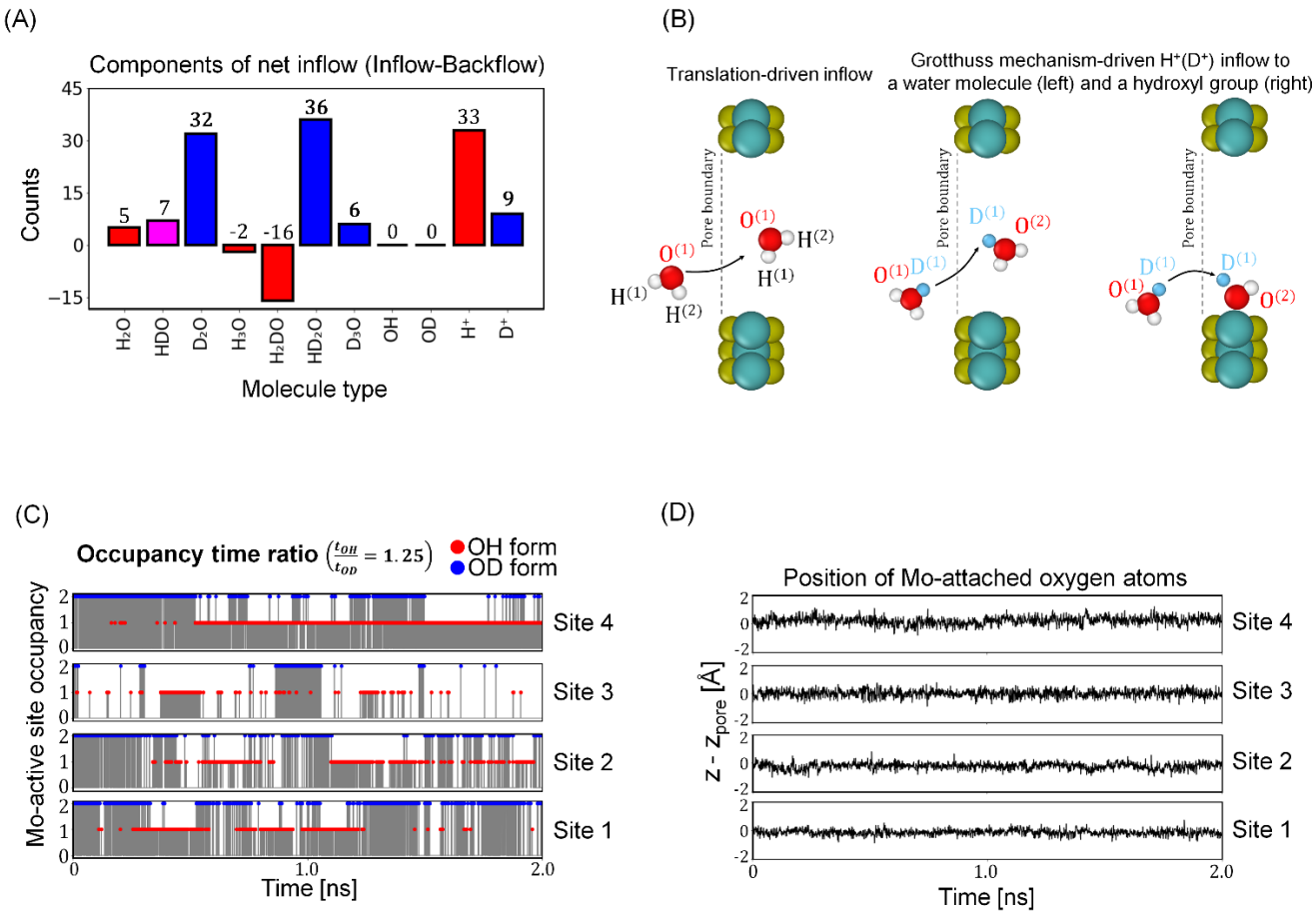

**Figure 3.** (A) The net inflow of molecules entering the pore (the number of molecules leaving the pore (backflow) is subtracted from the number entering the pore (inflow). The plots for inflow, backflow, and translocation are shown in supporting information; see Figure 1(C) for details). $D_2O$, D-dominant hydronium isotopes, and $H^+$ dominate the net inflow molecular group. (B) The mechanisms governing the net inflow. Translation-driven inflow, where a molecule crosses the pore boundary without a chemical reaction – water isotope and hydronium isotope molecules rely on this mechanism. $H^+$ and $D^+$ ions generally cross the pore boundary through the Grotthuss mechanism, where their acceptors are a water isotope molecule or a hydroxyl group attached to the Mo atom. (C) The visualization of the chemistry of the hydroxyl group attached to the Mo atom over time. OH attachment is preferred over OD attachment, quantified by the attachment time ratio ($t_{OH}/t_{OD}$) of 1.25. (D) The position of Mo-attached oxygen atoms, showing that they are almost stationary. That is, once $H^+$ and $D^+$ ions are attached to a hydroxyl group, they are trapped until another $H^+$ or $D^+$ ion combines with the hydroxyl group.

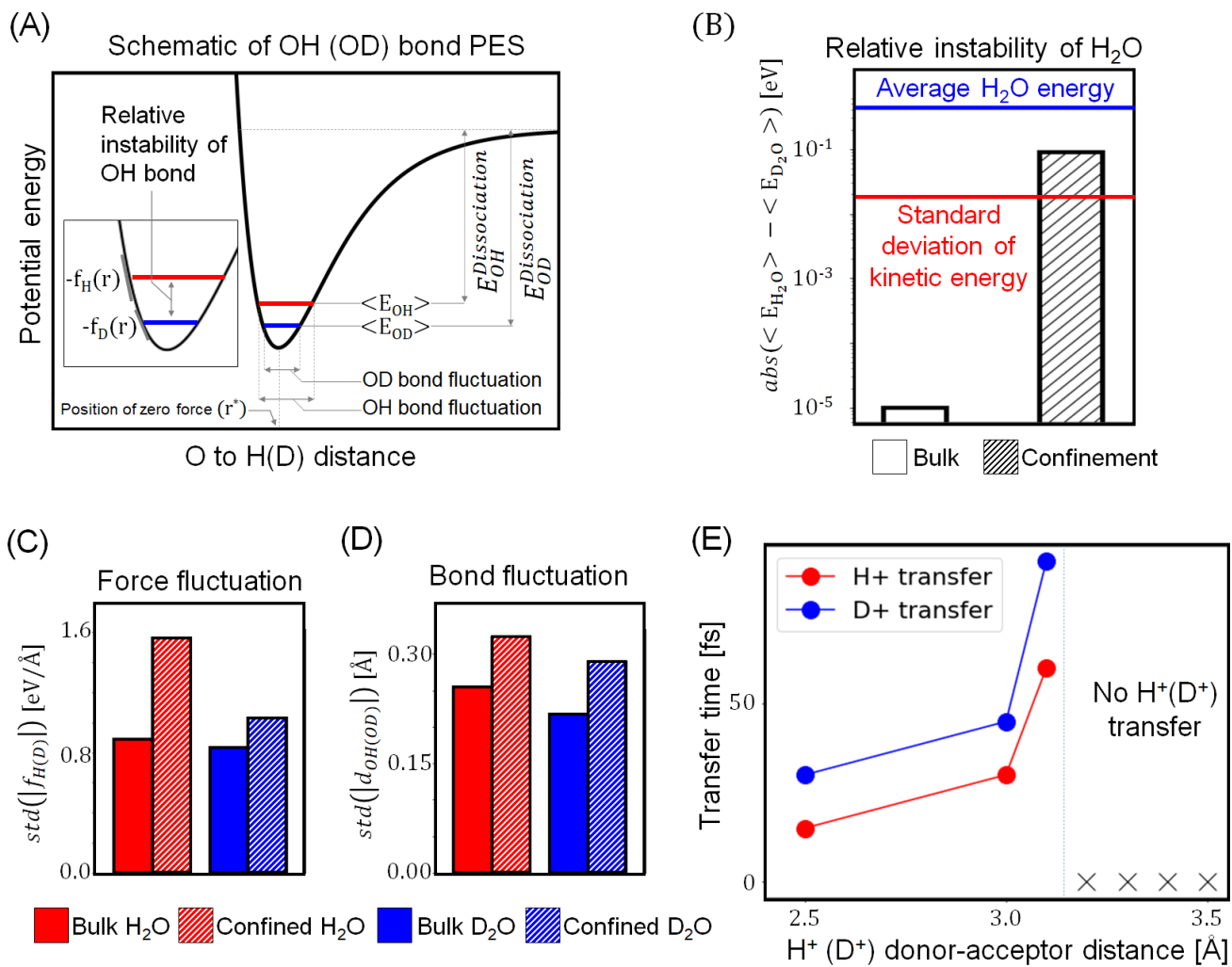

**Figure 4.** (A) Conceptual illustration of OH (OD) bond PES with quantities of interest: force, position of zero force ($r^*$), bond length fluctuation, average potential energy, and dissociation energy. The heavier mass of deuterium reduces the range of bond fluctuation, resulting in lower average potential energy and higher force fluctuation. The lower average energy of the OD bond requires more energy for dissociation, *i.e.*, higher dissociation energy. (B) The stability difference between water isotopes is quantified by the absolute value of average total energy difference. In bulk, the total energy of $H_2O$ and $D_2O$ are nearly the same, indicating they have similar stability. In confinement, the total energy of $H_2O$ is significantly higher than that of $D_2O$ implying lower stability of $H_2O$. To assess the significance of the energy difference, they are compared with the magnitude of average $H_2O$ energy (blue) and kinetic energy fluctuation (red). (C) The fluctuations of force acting on hydrogen and deuterium. Force fluctuations of $H_2O$ and $D_2O$ are higher in confinement than in bulk, indicating that both molecules are relatively unstable when they enter the pore. The increase in the force fluctuations for $H_2O$ is significantly larger compared to that of $D_2O$, implying $H_2O$ is relatively more unstable than $D_2O$. (D) The fluctuations in OH and OD bond lengths, which is another measure of stability. The bond fluctuations show a similar tendency to that of the force fluctuations: the fluctuations are smaller in bulk but larger in confinement, and $H_2O$ has higher fluctuation than $D_2O$ when they are in the same environment. (E) The time required

for a $H^+$ or $D^+$ ion transfer to a nearby Mo-attached oxygen atom. $H^+$ ion shows lower transfer time due to the higher mobility and the lower dissociation energy.